\documentclass[twocolumn]{IEEEtran}

\usepackage{amsmath,dsfont,bbm,epsfig,amssymb,amsfonts,amstext,verbatim,amsopn,cite}
\usepackage{subfigure,multirow,multicol,lipsum,xfrac}
\usepackage{amsthm,ulem}
\usepackage{mathtools,amsthm}
\usepackage{perpage}
\usepackage{balance}
\usepackage{url}
\usepackage{amsfonts}
\usepackage{epsfig}
\usepackage[font={small}]{caption}
\usepackage{etoolbox}
\usepackage{algorithmicx}
\usepackage[Algorithm,ruled]{algorithm}
\usepackage{algpseudocode}
\usepackage{pifont}
\usepackage[utf8]{inputenc}
\usepackage[T1]{fontenc}  
\usepackage[nolist]{acronym}
\MakePerPage{footnote}
\usepackage{textcomp}
\usepackage{paralist}
\usepackage{enumitem}
\usepackage{bbm}
\usepackage[process=auto]{pstool}

\newcommand{\setZ}{\mathbb{Z}}

\DeclareMathOperator*{\argmin}{\mathrm{argmin}\hspace*{1mm}}

\newcommand{\mOmega}{\mathbf{\Omega}}
\newcommand{\mB}{\mathbf{B}}
\newcommand{\mW}{\mathbf{W}}

\newcommand{\mT}{\mathbf{T}}

\newcommand{\R}{\mathbb{R}}
\newcommand{\Z}{\mathbb{Z}}

\newtheoremstyle{mystyle}
  {}
  {}
  {}
  {}
  {\bfseries}
  {:}
  { }
  {}
\theoremstyle{mystyle}

\newtheorem{definition}{Definition}

\newtheorem{example}{Example}
\algnewcommand\algorithmicLet{\textbf{Let}}
\algnewcommand\Let{\item[\algorithmicLet]}
\algnewcommand\algorithmicSet{\textbf{Set}}
\algnewcommand\Set{\item[\algorithmicSet]}

\algnewcommand\algorithmicInitiate{\textbf{Initiate}}
\algnewcommand\Initiate{\item[\algorithmicInitiate]}
\algnewcommand\algorithmicStart{\textbf{Begin}}
\algnewcommand\Begin{\item[\algorithmicStart]}
\algnewcommand\algorithmicEnd{\textbf{End}}
\algnewcommand\End{\item[\algorithmicEnd]}

\algnewcommand\algorithmicOutP{\textbf{Output:}}
\algnewcommand\Out{\item[\algorithmicOutP]}

\algnewcommand\algorithmicInP{\textbf{Input:}}
\algnewcommand\In{\item[\algorithmicInP]}

\newcounter{bar}

\renewcommand{\emph}[1]{\textit{#1}}
\def\expect{\mathop{\mbox{$\mathsf{E}$}}}

\begin{document}
\title{Linear Computation Coding}
\author{
\IEEEauthorblockN{
Ralf R. M\"uller\IEEEauthorrefmark{1},
Bernhard G\"ade\IEEEauthorrefmark{1},
Ali Bereyhi\IEEEauthorrefmark{1}\\
}
\IEEEauthorblockA{
\IEEEauthorrefmark{1}Institute for Digital Communications, Friedrich-Alexander Universit\"at Erlangen-N\"urnberg, Germany\\
ralf.r.mueller@fau.de, bernhard.gaede@fau.de, ali.bereyhi@fau.de
}
\thanks{This paper was/will be presented in part at the Information Theory \& Applications Workshop, San Diego, CA, Feb.\ 2020 and the IEEE Conference on Acoustics, Speech, and Signal Processing, Toronto, Canada, Jun.\ 2021.}
\thanks{This work was partially supported by Deutsche Forschungsgemeinschaft (DFG) under grant MU-3735/6-1.}
}
\IEEEoverridecommandlockouts
\maketitle

\begin{acronym}
\acro{oas}[OAS]{oversampled adaptive sensing}
\acro{csi}[CSI]{channel state information}
\acro{awgn}[AWGN]{additive white Gaussian noise}
\acro{iid}[i.i.d.]{independent and identically distributed}
\acro{rhs}[r.h.s.]{right hand side}
\acro{lhs}[l.h.s.]{left hand side}
\acro{wrt}[w.r.t.]{with respect to}
\acro{rs}[RS]{replica symmetry}
\acro{rsb}[RSB]{replica symmetry breaking}
\acro{mse}[MSE]{mean squared error}
\acro{mmse}[MMSE]{minimum MSE}
\acro{sinr}[SINR]{signal to interference and noise ratio}
\acro{mf}[MF]{matched filtering}
\end{acronym}
\begin{abstract}

We introduce the new concept of computation coding. Similar to how rate-distortion theory is concerned with the lossy compression of data, computation coding deals with the lossy computation of functions.

Particularizing to linear functions, we present an algorithm to reduce the computational cost of multiplying an arbitrary given matrix with an unknown column vector.
The algorithm decomposes the given matrix into the product of \textbf{\textit{codebook}} and \textbf{\textit{wiring}} matrices whose entries are either zero or signed integer powers of two.

For a typical implementation of deep neural networks, the proposed algorithm reduces the number of required addition units several times. To achieve the accuracy of 16-bit signed integer arithmetic for 4k-vectors, no multipliers and only 1.5 adders per matrix entry are needed. 
\end{abstract}

\IEEEpeerreviewmaketitle


\section{Motivation}
\label{sec:intro}
Neural networks are becoming an integral part of modern day's reality. This technology consists of two stages: A training phase and an inference phase. The training phase is computationally expensive and typically outsourced to cluster or cloud computing. It takes place only now and then, eventually only once forever. The inference phase is implemented on the device running the application. It is repeated whenever the neural network is used. This work solely targets the inference phase after the neural network has been successfully trained.

The inference phase consists of scalar nonlinearities and matrix-vector multiplications. The computational burden is overwhelmingly at the latter. The target of this work is to reduce the computational cost of the following task: Multiply an arbitrary unknown vector with a known, but arbitrary matrix. At the first layer of the neural network, the unknown vector is the input to the neural network. At a subsequent layer, it is the output of the respective previous layer. The known matrices are the outcome of the training phase in the data fusion center and fixed for all inference cycles of the neural network.

The computing unit running the inference phase need not be a general-purpose processor. With neural networks being more and more frequently deployed in low-energy devices, it is attractive to employ dedicated hardware. For some of them, e.g, field programmable gate arrays or application-specific integrated circuits with memory, the data center has the option to update the known matrices, whenever it wants to reconfigure the neural network. Still, the matrices stay fixed for most of the time. In this work, we will not address those updates, but focus on the most computationally costly effort: The frequent matrix-vector multiplications within the dedicated hardware.

Besides the matrix-vector multiplications, memory access is currently also considered a major bottleneck in the inference phase of neural networks. However, technological solutions to the memory access problem, e.g., stacked dynamical random access memory utilizing through-silicon vias \cite{shen:18} or emerging non-volatile memories \cite{hong:14}, are being developed and expected to be available soon. Thus, we will not address memory-access issues in this work. 

Various works have addressed the problem of simplifying matrix-matrix multiplications utilizing certain recursions that result in sub-cubic time-complexity of matrix-matrix multiplication (and matrix inversion) \cite{strassen:69,copperfield:90}. However, these algorithms and their more recent improvements, to the best of our knowledge, do not help for matrix-vector products. This work is not related to that group of ideas.

Various other works have addressed the problem of simplifying matrix-vector multiplications in neural networks utilizing structures of the matrices, e.g., sparsity \cite{han:15,louizos:17}. However, this approach comes with severe drawbacks: 1) It does not allow to design training phase and inference phase independently from each other. This restricts interoperability, hinders efficient training, and compromises performance \cite{evci:19}. 2) Sparsity alone does not necessarily reduce computational cost, as it may require higher accuracy, i.e. larger word-length for the nonzero matrix elements. In this work, we will neither utilize structures of the trained matrices nor structures of the input data. The vector and matrix to be multiplied may be totally arbitrary. They may, but need not, contain independent identically distributed (IID) random variables, for instance.

It is not obvious that, without any specific structure in the matrix, significant computational savings are possible over state-of-the-art methods implementing matrix-vector multiplications. In the sequel, we will explain why such savings are possible and how they can be achieved. We also show that these savings are very significant for typical matrix-sizes in present day's deep networks: By means of linear computation coding, the computational cost, if measured in number of additions and bit shifts, is reduced several times.

The paper is organized as follows:
In Section~\ref{gencomcod}, the general concept of computation coding is introduced.
In Section~\ref{staofart}, we review the state-of-the-art and define a benchmark for comparison.
In Section~\ref{proalg}, we propose our new algorithm.
Sections~\ref{perana} and \ref{simres} study its performance by analytic and simulative means, respectively.
Section~\ref{conc} summarizes conclusions and gives an outlook to open problems and applications of linear computation coding beyond the area of neural networks. 

Matrices are denoted by boldface upper letters, vectors are not explicitly distinguished from scalar variables. The sets $\Z$ and $\R$ denote the integers and reals, respectively. The identity matrix, the all zero matrix, the all one matrix, the expectation operator, the sign function, Landau's big O-operator, and the Frobenius norm are denoted by $\mathbf I$, $\mathbf 0$, $\mathbf 1$, $\expect [\cdot]$, $\text{sign}(\cdot)$, ${\mathsf O}(\cdot)$, and $||\cdot||_{\text F}$, respectively. Indices to constant matrices express their dimensions. The notation $||\cdot||_0$ counts the number of non-zero entries of the vector- or matrix-valued argument.
\section{Computation Coding for General Functions}
\label{gencomcod}

The approximation by a deep network is the current state-of-the-art to compute a multi-dimensional function efficiently. There may be other ones, yet undiscovered, as well. Thus, we define computation coding for general multi-dimensional functions. Subsequently, we discuss the practically important case of linear functions, i.e., matrix-vector products, in greater detail.

The best starting point to understand computation coding is rate-distortion theory in data compression. In fact, computation coding can be interpreted as a lossy encoding of functions with a side constraint on the computational cost of the decoding algorithm. It shares a common principle with lossy source coding: Random codebooks, if suitably constructed, usually perform well. In contrast to rate-distortion theory, however, random codebooks turn out useful even for practical applications of computation coding.

{\it Computation coding} consists of {\it computation encoding} and {\it computation decoding}. Roughly speaking, computation encoding is to find some approximate representation $m(x)$ for a given and known function $f(x)$ such that $m(x)$ can be calculated for any arbitrary $x\in {\cal X}$ with low computational cost and high accuracy. Computation decoding is the calculation of $m(x)$. Formal definitions are as follows:

\begin{definition}
\label{def1}
Given a probability space $({\cal X},P_{\cal X})$ and a metric $d:{\cal F} \times {\cal F}\mapsto \R$, a computation encoding with distortion $D$ for given function $f:{\cal X}\mapsto {\cal F}$ is a mapping $m:{\cal X} \mapsto {\cal M\subseteq F}$ such that $\expect_{x\in {\cal X}} [d(f(x),m(x))]\le D$.
\end{definition}
\begin{definition}
\label{def2}
A computation decoding with computational cost $C$ for given operator $\mathsf C$ is an implementation of the mapping $m:{\cal X}\mapsto {\cal M}$ such that $\mathsf C[m(x)]\le C$ for all $x\in{\cal X}$.
\end{definition}

The computational cost operator $\mathsf C[m(\cdot)]$ measures the cost to implement the function $m(\cdot)$. It reflects the properties of the hardware that executes the computation. 

Computation coding can be regarded as a generalization of lossy source coding.
If we consider the identity function $f(x)=x$ and the limit $C\to\infty$, computation coding specializes to lossy source coding with $m(x)$ being the lossy codeword for $x$.
Rate-distortion theory analyzes the trade-off between distortion $D$ and the code rate. In computation coding, we are interested in the trade-off between distortion $D$ and computational cost $C$. The code rate is of no or at most subordinate concern.

The expectation operator in the distortion constraint of Definition~\ref{def1} is natural to readers familiar with rate-distortion theory. From a computer science perspective, it follows the philosophy of approximate computing \cite{palem:13}. Nevertheless, hard constraints on the accuracy of computation can be addressed via distortion metrics based on the infinity norm, which enforces a maximum tolerable distortion.

The computational cost operator may also include an expectation. Whether this is appropriate or not, depends on the goal of the hardware design. If the purpose is minimum chip area, one usually must be able to deal with the worst case and an expectation can be inappropriate. Power consumption, on the other hand, overwhelmingly correlates with average computational cost.

The above definitions shall not be confused with related, but different definitions in the literature of approximation theory \cite{devore:98}. There, the purpose is rather to allow for proving theoretical achievability bounds than evaluating algorithms. The approach to distortion is similar. Complexity, however, is measured as the growth rate of the number of bits required to achieve a given upper bound on distortion. This is quite different from the computational cost in Definition~\ref{def2}.   

\section{State of the Art}
\label{staofart}

\subsection{Scalar Linear Functions}
\label{lincomcod}

The principles and benefits of computation coding are most easily explained for linear functions. Let us start with scalar linear functions, before we get to the multi-dimensional case.

\begin{example}
\label{ex1}
Let $x\in\R$ have unit variance and $f(x)=tx,t\in\R$. Let the computation encoding take the form
\begin{align} 
m(x)= {\rm sign}(t) \sum\limits_{b\in{\cal B}}2^b x
\end{align}
where the set ${\cal B}\subset \Z$ simply contains the positions of all the 1-bits in the appropriately rounded binary representation of $t$. Define the computational cost operator as ${\mathsf C}[m(x)]=\max {\cal B} -\min {\cal B} +1$, independent of $x$. This is the standard way a linear function is implemented on a modern computer by means of additions and bit shifts. Considering $t$ as a random variable with uniform distribution on $[-1,+1]$, the trade-off between average mean-squared distortion and computational cost is well-known to be \cite{gray:98}
\begin{align}
\expect\limits_{x\in\R} [f(x)-m(x)]^2=4^{-C}/3,
\label{ssia}
\end{align}
i.e., every additional bit of resolution reduces the quantization error by 6 dB. 
\end{example}

On average, half of the additions are actually additions of zero. For 16-bit signed integer arithmetic, we only need to add $(16-1)/2=7.5$ terms, on average, thus, compute 6.5 additions. The multiplication of a $512\times 4096$ matrix of 16-bit signed fixed-point numbers by an unknown column vector, thus, requires $512\cdot(4096\cdot 6.5+4095)\approx 15.7$ million additions, i.e., 7.5 additions per matrix entry, on average. It achieves an average distortion of $4^{-15}/3$ which is equivalent to $-95$ dB. This is used as benchmark in the sequel.

For very high precision, i.e. very small distortion, more efficient algorithms are known from literature that are all based on some version of the Chinese Remainder Theorem \cite{karatsuba:62,toom:63,cook:69,schoenhage:71,fuerer:09,harvey:16}. For precisions relevant in neural networks, however, these algorithms are not useful.

\begin{example}
\label{ex2}
Consider the setting of Example~\ref{ex1}, except for the computation encoding to take the canonical signed digit form \cite{booth:51}
\begin{align}
m(x)= \sum\limits_{(s,b)\in{\cal P}}s2^b x 
\end{align}
with ${\cal P} \subset \{\pm 1\}\times \Z$. Like in Example~\ref{ex1}, the number $t$ is approximated by a weighted sum of powers of 2. However, the weighting coefficients are chosen from the set $\{-1,0,+1\}$ instead of $\{0,1\}$. The computational cost operator is defined as ${\mathsf C}[m(x)]=|{\cal P}|$. In contrast to Example~\ref{ex1}, it does not impose any cost for additions of zeros, but only counts the number of additions and subtractions. For the optimum choice of the set ${\cal P}$, the trade-off between average mean-squared distortion and computational cost is shown in the appendix to be
\begin{align}
\expect\limits_{x\in\R} [f(x)-m(x)]^2 =28^{-C}/3.
\end{align}
Thus, every unit increment of the computational cost reduces the quantization error by 14.5 dB. In this case, the average distortion of $-95$ dB is achieved for an average of $C=\log_{28} 4^{15}\approx 6.24$ additions per matrix entry. That is a 17\% reduction over the benchmark in Example~\ref{ex1}.
\end{example}

There are various further improvements to computation coding of scalar linear functions in the literature, e.g., redundant logarithmic arithmetic \cite{arnold:90,huang:94}.  One can also search for repeating bit patterns in the representation of $t$ and reuse the initial computation for later occurrences 
\cite{hartley:96, lefevre:00}. One can represent $t$ by multiple bases at the same time \cite{dimitrov:99}. However, the conversion into the multiple base number system does not come for free. Area and power savings were studied in \cite{dimitrov:11} for 32-bit and 64-bit integer arithmetic. Only for 64-bit arithmetic improvements were reported.

\subsection{Multidimensional Linear Functions}
\label{mullinfun}

Linear computation coding for products of structured matrices with unknown real or complex vectors is widely spread in the literature. The most known example is probably the fast Fourier transform. By contrast, computation coding for linear functions with unstructured matrices has rarely been studied at all with the notable exception of \cite{boullis:05,aksoy:16}. Reference \cite{boullis:05} finds up to 40\% improvements for structured matrices, but the performance for random matrices is reported as only ``slightly better''. In \cite{aksoy:16}, the processing on the matrix entries is optimized utilizing various algorithms including linear programming and pattern search. Computational cost for random matrices is reported to be reduced by at most 28\% over the scalar canonically signed digit form.

Accelerations to products of unstructured matrices with unknown vectors is well-studied for Boolean semi-rings \cite{kronrod:70,williams:07}. For Boolean semi-rings, \cite{williams:07} shows that lossless linear computation coding in $K$ dimensions requires at most ${\mathsf O}(K^2(\log K)^{-2})$ operations.
The mailman algorithm \cite{liberty:09} is inspired by these ideas.
It allows to compute the product of a matrix composed of entries from a finite field with an arbitrary (even real or complex) vector by at most ${\mathsf O}(K^2(\log K)^{-1})$ operations.
For this purpose, the target matrix 
\begin{equation}
\mT = \mB\mW
\label{decomp}
\end{equation}
is decomposed into a codebook matrix $\mB$ and a wiring matrix $\mW$ in order to simplify the computation of the linear function $f(x) = \mT x$ by $f(x) = \mB(\mW x)$ with appropriate choices of the codebook and the wiring matrix.

Let $\mT\in {\cal T}^{N\times K}$ for the finite set ${\cal T}=\{0,\dots,T-1\}$ with cardinality $T$.
Let $K=T^N$ and fix the codebook matrix $\mB \in {\cal T}^{N\times K}$ in such a way that entry $\mB_{nk}$ is the $n^\text{th}$ digit of the $T$-ary representation of $k-1$. Thus, the $k^{\text{th}}$ column of the codebook matrix is the $T$-ary representation of $k-1$.
Since $\mB$ contains all the $T^N$ possible columns, any column of $\mT$ is also a column of $\mB$.
The purpose of the wiring matrix is to pick the columns of the codebook matrix in the right order similar to a mailman who orders the letters suitably prior to delivery.
Since the wiring matrix only reorders the columns of the codebook matrix, it contains a single one per column while all other elements are zero. Thus, the product $h = \mW x$ does not require any arithmetic. On a circuit board, it just defines a wiring. For the product $\mB h$, \cite{liberty:09} gives a recursive algorithm that only requires ${\mathsf O}(K)$ operations.
Decomposing a $K\times K$ matrix into $K/\log_T K$ submatrices of size $\log_TK\times K$, the overall complexity scales as ${\mathsf O}(K^2/\log K)$.

The drawback of the mailman algorithm is that it can be used only for very low accuracy of computation. To reach the accuracy of $q$-bit unsigned integer arithmetic, we need matrices whose size is at least $2^q \times 2^{(2^q)}$. Arbitrary accuracy can only be achieved for infinite matrix size. Furthermore, the matrix size must grow exponentially with the desired signal-to-quantization-noise-ratio.

\section{Proposed Algorithm}
\label{proalg}

We start with the decomposition of the target matrix into codebook matrix and wiring matrix as in \eqref{decomp}. However, we design at least the wiring matrix in a manner different from \cite{liberty:09}.
We also drop the purity of wiring and allow some computational effort to calculate the product $\mW x$.

\subsection{Single Wiring Matrix}
\label{sinwirmat}

For given target matrix $\mT=[t_1,\dots,t_K] \in \R^{N\times K}$ and codebook matrix $\mB\in \{0, \pm 2^{\Z}\}^{N\times K}$, we find the wiring matrix $\mW=[w_1,\dots,w_K]$ such that it is a good solution to the sparse recovery problem
\begin{equation}
\label{comsenopt}
\mW = \argmin\limits_{\mOmega\in\{0,\pm 2^{\setZ}\}^{K\times K} :||\mOmega||_0=K+S} || \mT - \mB \mOmega ||_{\text F}
\end{equation}
for some parameter $S$ controlling the computational cost. 
This wiring matrix requires $S$ additions and $K+S$ shift operations.

Since the optimum solution to the optimization problem \eqref{comsenopt} is hard to find in polynomial time (NP-hard), we propose a greedy algorithm that is demonstrated in Sections~\ref{perana} and \ref{simres} to perform well.
First, we break the matrix optimization problem into $K$ column-wise optimization problems:
\begin{equation}
\label{comsencol}
w_k = \argmin\limits_{\omega\in\{0,\pm 2^{\setZ}\}^K :||\omega||_0=1+s} || t_k - \mB \omega ||_2 \qquad \forall k
\end{equation}
with $s=S/K$.
Since the column-wise optimization \eqref{comsencol} is still NP-hard, we take the following greedy approach for each column:
\begin{enumerate}
\item
Start with $s=0$ and $\omega=\mathbf 0_{N\times 1}$.
\item
Update $\omega$ such that it changes in at most a single component. 
\item
Increment $s$.
\item
If $s\le S/K$, go to step 2.
\end{enumerate}

The codebook matrix need not be a binary mailman matrix. Any matrix that can be multiplied with little effort to the product $h=\mW x$ is welcome. Details are discussed in Section~\ref{codmat}.

\subsection{Multiple Wiring Matrices}

The wiring matrix can be further decomposed into $L$ sub-wiring matrices
\begin{equation}
\mW = \mW_1 \mW_2 \cdots \mW_L.
\end{equation}
This means that $\mB$ serves as codebook for $\mW_1$ and $\mB\mW_1\cdots \mW_\ell$ serves as codebook for $\mW_{\ell+1}$. Repeating the greedy algorithm of Section~\ref{sinwirmat} for $\ell$ increasing from $1$ to $L$, the wiring matrices $\mW_\ell$ are recursively found.

Multiple wiring matrices are useful, if the codebook matrix $\mB$ is computationally cheap, but poor from a distortion point of view. The product of a computationally cheap codebook matrix $\mB$ with a computationally cheap wiring matrix $\mW_1$ can serve as a codebook $\mB\mW_1$ for subsequent wiring matrices that performs well with respect to both distortion {\it and} computational cost.

Multiple wiring matrices can also be useful if the hardware in the inference phase favors some serial over fully parallel processing. In this case, circuitry for multiplying with $\mW_\ell$ can be reused for subsequent multiplication with $\mW_{\ell-1}$. Note that in the design phase, wiring matrices are preferably calculated in increasing order of the index $\ell$, while in the inference phase, they are used in decreasing order of $\ell$.

\subsection{Codebook Matrices}
\label{codmat}

For codebook matrices, the computational cost depends on the way they are designed. Besides being easy to multiply onto a given vector, a codebook matrix should be designed such that pairs of columns are not collinear. A column that is collinear to another one is obsolete: It does not help to reduce the distortion while it requires to compute additions. In an early conference version of this work \cite{mueller:20}, we proposed to find the codebook matrix by sparse quantization of the target matrix.  While this results in significant savings of computational cost over the state of the art, there are even better designs for codebook matrices. Three of them are detailed in the sequel.

\subsubsection{Binary Mailman Codebook}
In the binary mailman codebook, only the all zero column is obsolete. It is shown in the appendix that the multiplication of the binary mailman matrix with an arbitrary vector requires less than $2K$ additions. While performing well, it lacks flexibility, as it requires the matrix dimensions to fulfill $K=2^N$. 

\subsubsection{Two-Sparse Codebook}
We choose the alphabet ${\cal S} \subset \{ 0, \pm 2^0, \pm 2^1, \pm 2^2, \dots \}$ as a subset of the signed positive powers of two augmented by zero and find $K$ vectors of lengths $N$ such that no pair of vectors is collinear and each vector has zero norm equal to either 1 or 2. For sufficiently large size of the subset, those vectors always exist. These vectors are the columns of the codebook matrix. The ordering is irrelevant. It turns out useful to restrict the magnitude of the elements of ${\cal S}$ to the minimum that is required to avoid collinear pairs. 

\subsubsection{Self-Designing Codebook}
We set $\mB = \mB_0 \mB_1 \mB_2$ with $\mB_0 = [\mathbf I_N \ \mathbf 0_{N\times (K-N)}]$ and find the $K\times K$ matrices $\mB_1$ and $\mB_2$ recursively via \eqref{comsenopt} for $S=K$ interpreting them as wiring matrices for some given auxiliary target matrix $\tilde{\mT}$. The auxiliary target matrix may, but need not be identical to $\mT$.
The codebook designs itself taking the auxiliary target matrix as model.

\section{Performance Analysis}
\label{perana}

In order to analyze the expected distortion, we resort to IID Gaussian random codebooks and target vectors, as well as mean-square distortion averaged over the codebook ensemble.
The IID Gaussian codebook is solely chosen, as it simplifies the performance analysis. In practice, the IID Gaussian random matrix $\mB$ must be replaced by a codebook matrix with low computational cost, but similar performance. Simulation results in Section~\ref{simres} will show that practical codebooks perform very similar to IID Gaussian ones.

\subsection{Logarithmic Aspect Ratio}
\label{logasprat}
A key point to good performance of the multiplicative matrix decomposition \eqref{decomp} is the logrithmic aspect ratio. The number of rows of the codebook matrix $N$ scales logarithmically with the number of columns $K$. For a linear computation code, we define the code rate as 
\begin{equation}
R = \frac 1N \log_2 K.
\end{equation}
The code rate is a design parameter that, as we will see later on, has some minor impact on the trade-off between distortion and computational cost.

\begin{figure}
\centerline{
\epsfig{file=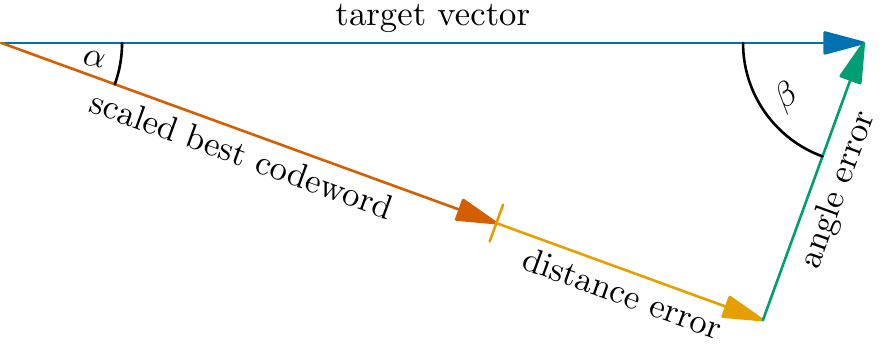,width=.9\columnwidth}}
\caption{\label{pic} Decomposition of the approximation error.}
\end{figure}

The logarithmic scaling of the aspect ratio is fundamental.
This is a consequence of extreme-value statistics of large-dimensional random vectors: Consider the correlation coefficients (inner products normalized by their norms) of $N$-dimensional real random vectors with IID entries in the limit $N\to\infty$. For any set of those vectors whose size is polynomial in $N$, the squared maximum of all correlation coefficients converges to zero, as $N\to\infty$ \cite{jiang:04}. Thus, the angle $\alpha$ in Fig.~\ref{pic} 
becomes a right angle and the norm of the angle error is lower bounded by the norm of the target vector. However, for an exponentially large set of size $2^{RN}$ with rate $R>0$, the limit for $N\to\infty$ is strictly positive and given by rate-distortion theory as $1-4^{-R}$ \cite{berger:71}. The asymptotic squared relative error of approximating a target vector by an optimal scaling of the best codeword, thus, is $4^{-R}$. The error itself can be approximated by another vector of the exponentially large set to get the total squared error down to $4^{-2R}$. Repeating that procedure for $s$ times, the (squared) error decays exponentially in $s$. In practice, the scale factor  cannot be a real number, but must be quantized. This additional error is illustrated in Fig.~\ref{pic} and labelled distance error as opposed to the previously discussed angle error. As a result, the total squared error does not decay as $4^{-(s+1) R}$, but with a slightly reduced rate.

The logarithmic aspect ratio has the following impact on the trade-off between distortion and computational cost:
For $K+S$ choices from the codebook, the wiring matrix has $K+S$ nonzero entries. This implies $S$ additions in the product $h= \mW x$.  
In return, we get approximated  an $N\times K$ target matrix $\mT$ with $N=\frac1R\log_2 K = \mathsf O(\log K)$ rows. For any desired distortion $D$, the computational cost of the product $\mW x$ is by a factor $\mathsf O(\log K)$ smaller than the number of entries in the target matrix. This is the same scaling as in the mailman algorithm. 
Given such a scaling, the mailman algorithm allows for a fixed distortion $D$ which depends on the size of the target matrix. 
The proposed algorithm, however, can achieve arbitrarily low distortion by setting $S$ appropriately large irrespective of the matrix size. The connection between the computational cost $C=2K+S$ and distortion $D$ is addressed in the sequel.

\subsection{Angle Error}

Consider a unit norm target vector $t\in\R^{N}$ that shall be approximated by a scaled version of one out of $K$ codewords $b_k\in\R^{N}$ which are random and jointly independent.
Denoting the angle between the target vector $t$ and the codeword $b_k$ as $\alpha_k$, we get (norm of) the angle error as
\begin{equation}
a_k = |\sin \alpha_k|.
\end{equation}
The correlation coefficient between target vector $t$ and codeword $b_k$ is given as 
\begin{equation}
\rho_k = \frac{<t;b_k>}{||t||_2\, ||b_k||_2} = \cos\alpha_k
\end{equation}
with $<\cdot;\cdot\cdot>$ denoting the inner product.
The minimum angle error $a$ and the correlation coefficients are related by 
\begin{equation}
a^2= 1 - \max\limits_k \rho_k^2.
\end{equation}
Next, we will study the behavior of the correlation coefficients in order to learn about the minimum angle error.

Let ${\text P}_{\rho^2|t}(r,t)$ denote the cumulative distribution function (CDF) of the squared correlation coefficient given target vector $t$. 
As the columns of the codebook matrix are jointly independent, we conclude that for
\begin{equation}
\varrho^2 = \max_k \rho_k^2,
\end{equation}
we have
\begin{equation}
{\text P}_{\varrho^2|t}(r,t) = \left[{\text P}_{\rho^2|t}(r,t)\right]^K.
\end{equation}

The target vector $t$ follows a unitarily invariant distribution. Thus, the conditional CDF does not depend on it. In the sequel, we choose $t$ to be the first unit vector of the coordinate system, without loss of generality.

\begin{figure}
\centerline{\epsfig{file=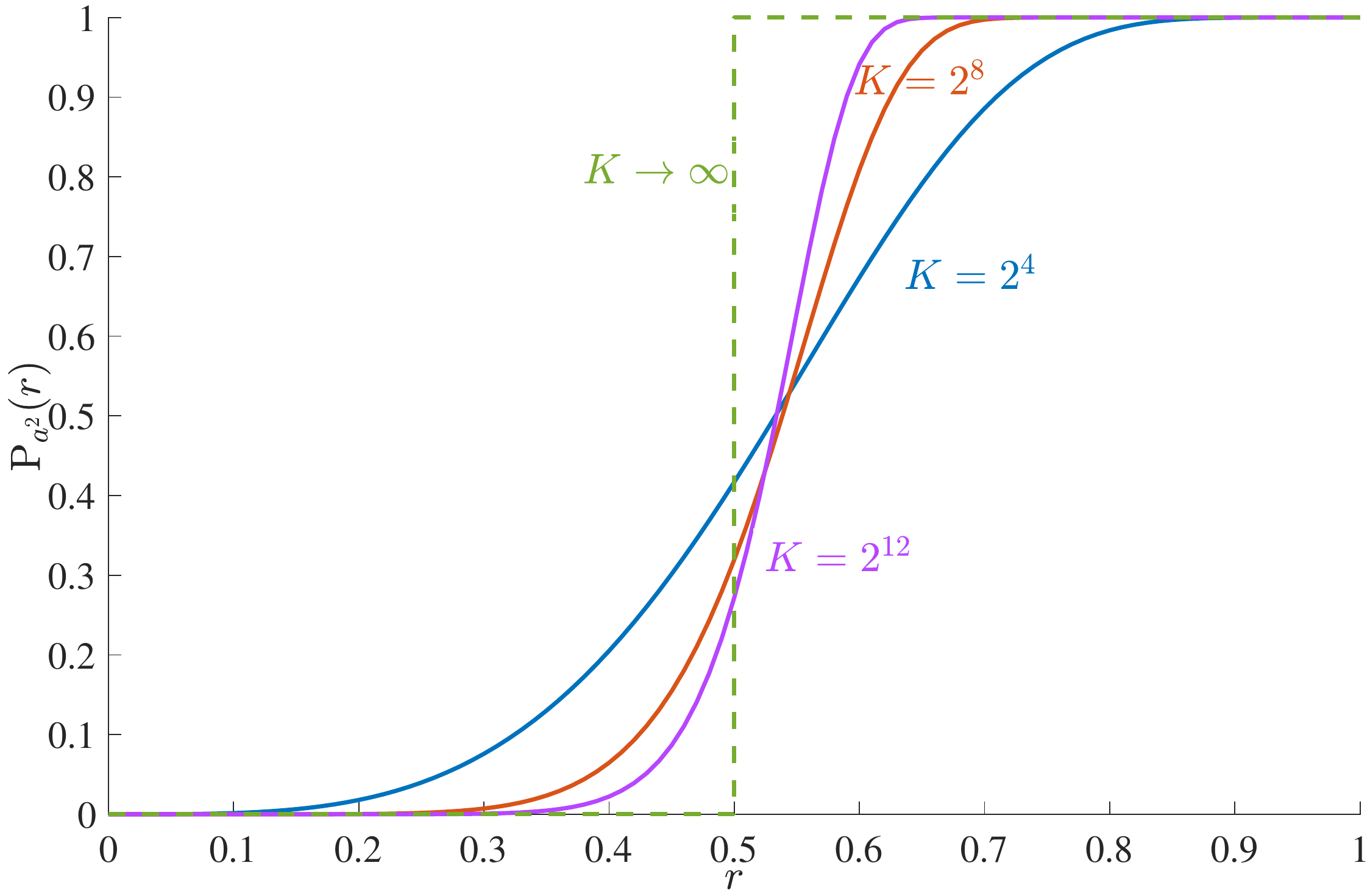,width=\columnwidth}}
\caption{CDF of the squared angle error for rate $R=\frac 12$ and various numbers of columns $K$.
\label{fig_cdf}}
\end{figure}

The squared correlation coefficient $\rho_k^2$ is known to be distributed according to the beta distribution with shape parameters $\frac12$ and $\frac {N-1}2$ \cite[Sec. III.A]{mueller:20a} and given by
\begin{equation}
{\text P}_{\rho^2}(r) = {\text B} \left(\frac12,\frac{N-1}2,r\right)
\end{equation}
with 
\begin{equation}
{\text B} (a,b,x) =  \frac{\int\limits_{0}^x \xi^{a-1} (1-\xi)^{b-1} {\text d}\xi}  {\int\limits_{0}^1 \xi^{a-1} (1-\xi)^{b-1} {\text d}\xi}
\end{equation}
denoting the regularized incomplete Beta function \cite{spanier:87}.
The distribution of the squared angle error is, thus, given by
\begin{align}
{\text P}_{a^2}(r) &= 1-\left[{\text B} \left(\frac12,\frac{N-1}2,1-r\right)\right]^K.
\end{align}
It is shown in the appendix to converge to
\begin{equation}
\lim\limits_{K\to\infty} \lim\limits_{N\to\frac1R\log_2 K} {\text P}_{a^2}(r) = \left\{
\begin{array}{ll}
0 & r < 4^{-R}\\
1 & r > 4^{-R}
\end{array}
\right..
\label{asymptoticdis}
\end{equation}
for logarithmic aspect ratios confirming the considerations in Section~\ref{logasprat}.
The CDF is depicted in Fig.~\ref{fig_cdf}.
For finite dimensions, we find the mean squared angle error as
\begin{align}
\overline{a^2} &= \int\limits_0^1 a\, \text{dP}_{a^2}(a) = 1- \int\limits_0^1 \text{P}_{a^2}(a) {\text d} a\\
& =  \int\limits_0^1 {\text B} \left(\frac12,\frac{N-1}2,r\right)^K {\text d}r.
\end{align}
using integration by parts.

\subsection{Distance Error}
\label{diserr}

For a target vector of unit norm, the distance error $d$ is bounded from above by 
$\frac 13 |\cos \alpha|$.
To see this, note that the norm of the optimally scaled codeword is $|\cos \alpha|$. Furthermore, the maximum error occurs, if the magnitude of the optimum scale factor $v$ is exactly in the middle of two adjacent powers of two, say $p$ and $2p$.
In this case, we have 
\[
d= |v| -p = 2p - |v| \qquad \Rightarrow \qquad |v|-p = \frac{|v|}3.
\]
Assuming the error to be uniformly 
distributed, it is easily concluded that the average squared distance error is given by 
\begin{equation}
\overline{d^2} = \frac1{27}\, \overline{\varrho^2} = \frac{1-\overline{a^2}}{27}.
\end{equation}

Note that the factor $1/27$ slightly differs from the factor $1/28$ in Example~\ref{ex2}. Like in Example~\ref{ex2}, the number to be quantized is uniformly distributed within some interval. Here however, the interval boundaries are not signed powers of two. This leads to a minor increase in the power of the quantization noise.

\subsection{Total Error}
\label{toterr}

\begin{figure}
\centerline{\epsfig{file=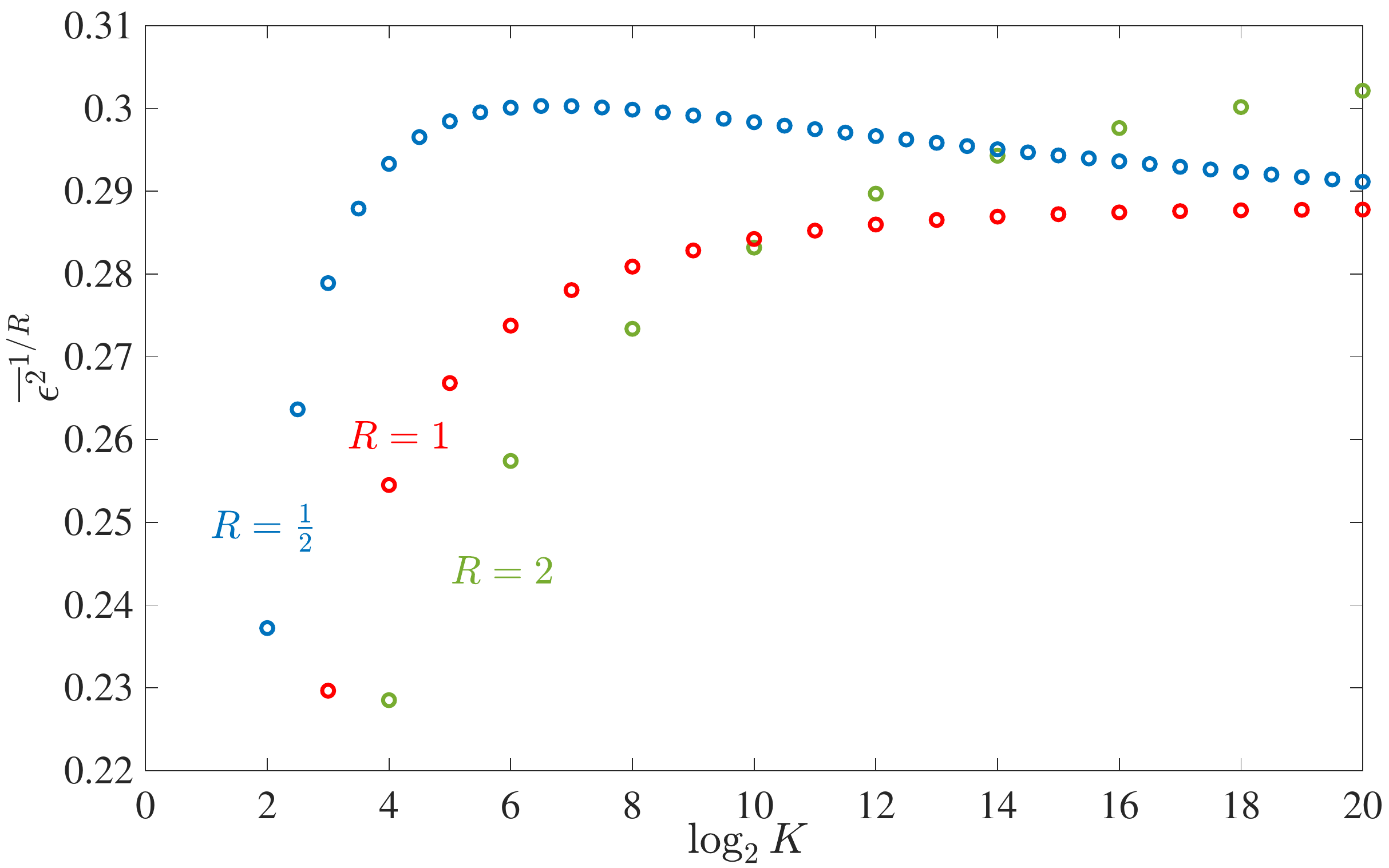,width=\columnwidth}}
\caption{$R^{\text{th}}$ root of the average total squared error vs.\ the number of columns $K$ in logarithmic scale for various rates $R$.
\label{totalerror}}
\end{figure}

Since distance error and angle error are orthogonal to each other, the average total squared error is simply given as
\begin{equation}
\overline{\epsilon^2} = \overline{a^2} + \overline{d^2} = \frac{1+26 \, \overline{a^2}}{27}.
\end{equation}
The average total squared error for a single approximation step is depicted in Fig.~\ref{totalerror} for various rates $R$.
Note that the computational cost per matrix entry linearly scales with $R$ for fixed $K$. In order to have a fair comparison, the average total squared error is exponentiated with $1/R$.
Despite the finite size of the matrices and the additional distance error due to quantization of the scale factor, it does not deviate very much from the asymptotic value of $\frac14$ found in \eqref{asymptoticdis}.

Every reduction of error compounds on top of the previous one. If the errors in $s+1$ subsequent steps of approximation were statistically independent, the remaining average total squared error would be 
\begin{equation}
\label{lowerbound}
D_{\text{LB}} = 
\left(\overline{\epsilon^2}\right)^{s+1}.
\end{equation}
for $s+1$ steps. As discussed in the next paragraph, they are not independent and \eqref{lowerbound} is just a lower bound.

The angle between the target vector and the best codeword (denoted by $\alpha$) and the angle between target vector and angle error (denoted by $\beta$) add to $90^\circ$, see Fig.~\ref{pic}.
The larger the angle error, the smaller the angle $\beta$. That means a large angle error implies that the next target vector is close to the previous one. In other words, a target vector that is difficult to approximate implies that the next target vector is likely to be difficult to approximate, as well. On the other hand, a target vector that is easy to approximate results in a new target vector that is close to orthogonal to the previous target vector. In this case, it is pretty much a fair coin flip to decide whether the next approximation is easy or difficult. Thus, it is more likely to get a difficult target vector than an easy one, on average.
This effect is all the more pronounced the smaller the size of the codebook $K$ is. This is because smaller values of $K$ increase the probability of having a large angle error, see Fig.~\ref{fig_cdf}.

\subsection{Computational Cost}
\label{comcos}

For sake of simplicity, we use the number of additions to measure computational cost. Sign changes and shifts are cheaper than additions and their numbers are also very close to the number of additions.

For wiring matrix $\mW_\ell$ with $1+s_\ell$ nonzero entries per column, we have to sum $K$ times $1+s_\ell$ terms. So, we need $s_\ell K$ additions in total for wiring matrix $\mW_\ell$. 
All three codebook matrices discussed in Section~\ref{codmat} require at most $2K$ additions.

Adding the computational costs of wiring and codebook matrices, we get
\begin{align}
C = 2K +\sum\limits_{\ell=1}^L s_\ell K = (s+2) K
\end{align}
with $s=\sum_{\ell=1}^L s_\ell$. Normalizing to the number of elements of the $N\times K$ target matrix $\mT$, we have
\begin{align}
\tilde C = \frac{C}{KN} = \frac{s+2}N.
\end{align}
The computational cost per matrix entry vanishes with increasing matrix size.
This behavior fundamentally differs from state-of-the-art methods discussed in Examples~\ref{ex1} and \ref{ex2}, where the matrix size has no impact on the computational cost per matrix entry.



\section{Simulation Results}
\label{simres}

\subsection{Accuracy of Lower Bound}
\label{acclowbou}

\begin{figure}
\epsfig{file=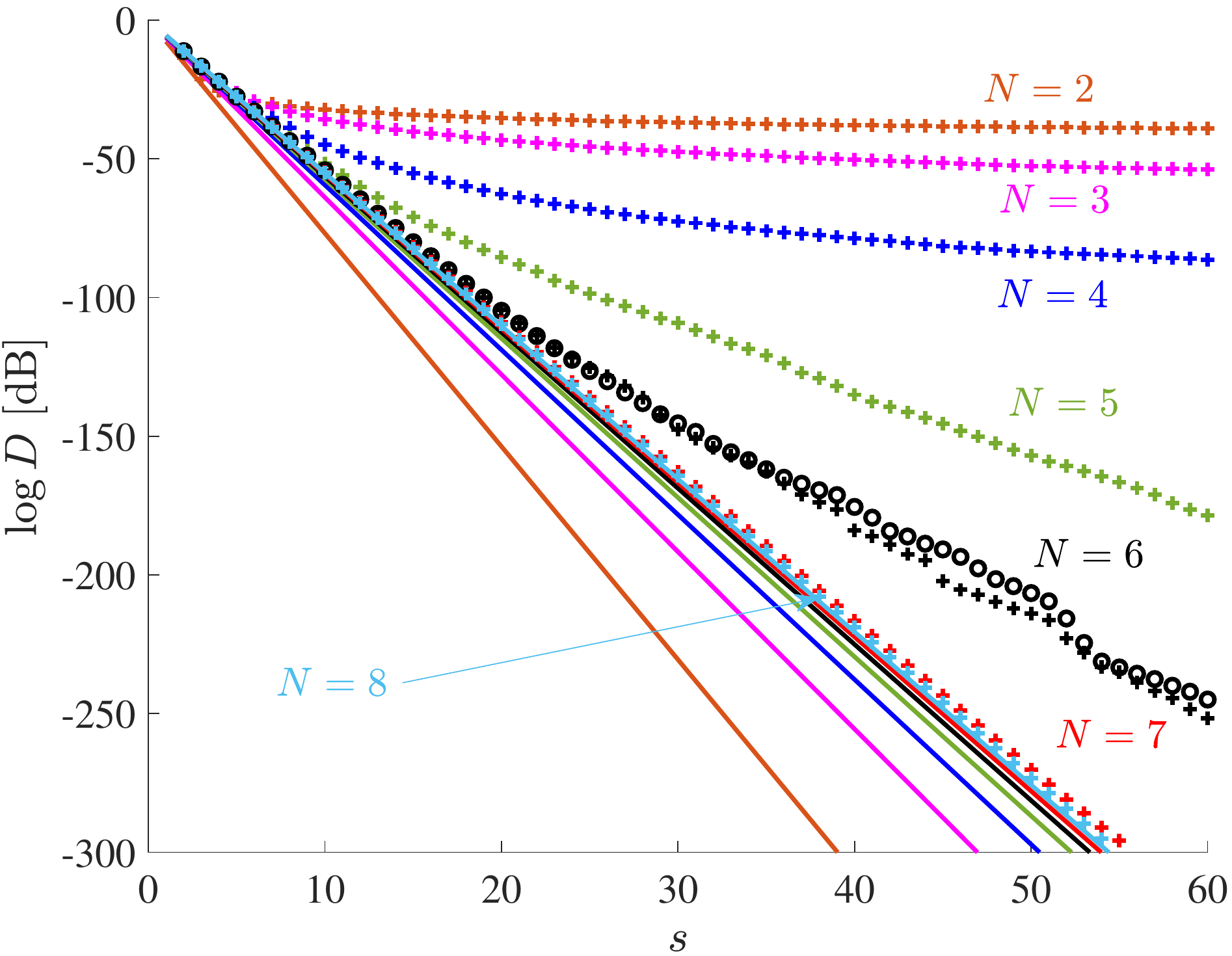,width=\columnwidth}
\caption{Average relative total squared error vs.\ total number of additions per column required for the wiring matrices. The solid lines refer to the lower bound \eqref{lowerbound}. The markers to simulation results. All results are for IID Gaussian codebook matrices with unit rate. Results are averaged over $10^6$ random matrix entries, except for the circular markers which refer to $10^7$ random matrix entries.
\label{testLB}}
\end{figure}

Consider $L$ wiring matrices each with two nonzero entries per column, i.e., $s_\ell=1$ for all $\ell$.
The total average distortion for IID Gaussian codebook matrices is shown in Fig.~\ref{testLB} 
vs.\ $s$, i.e., the total number of additions per column required for the wiring matrices.
Results are averaged over $10^6$ random matrix entries.
For unit rate, the lower bound \eqref{lowerbound} is very tight for $N>6$ and otherwise poor over a wide range of $s$.
Remarkably, the jittery behavior for $N=6$ seems not to result from insufficient averaging, as going for ten times more matrix samples does not reduce it. 

\subsection{Codebook Comparison}

\begin{figure}
\epsfig{file=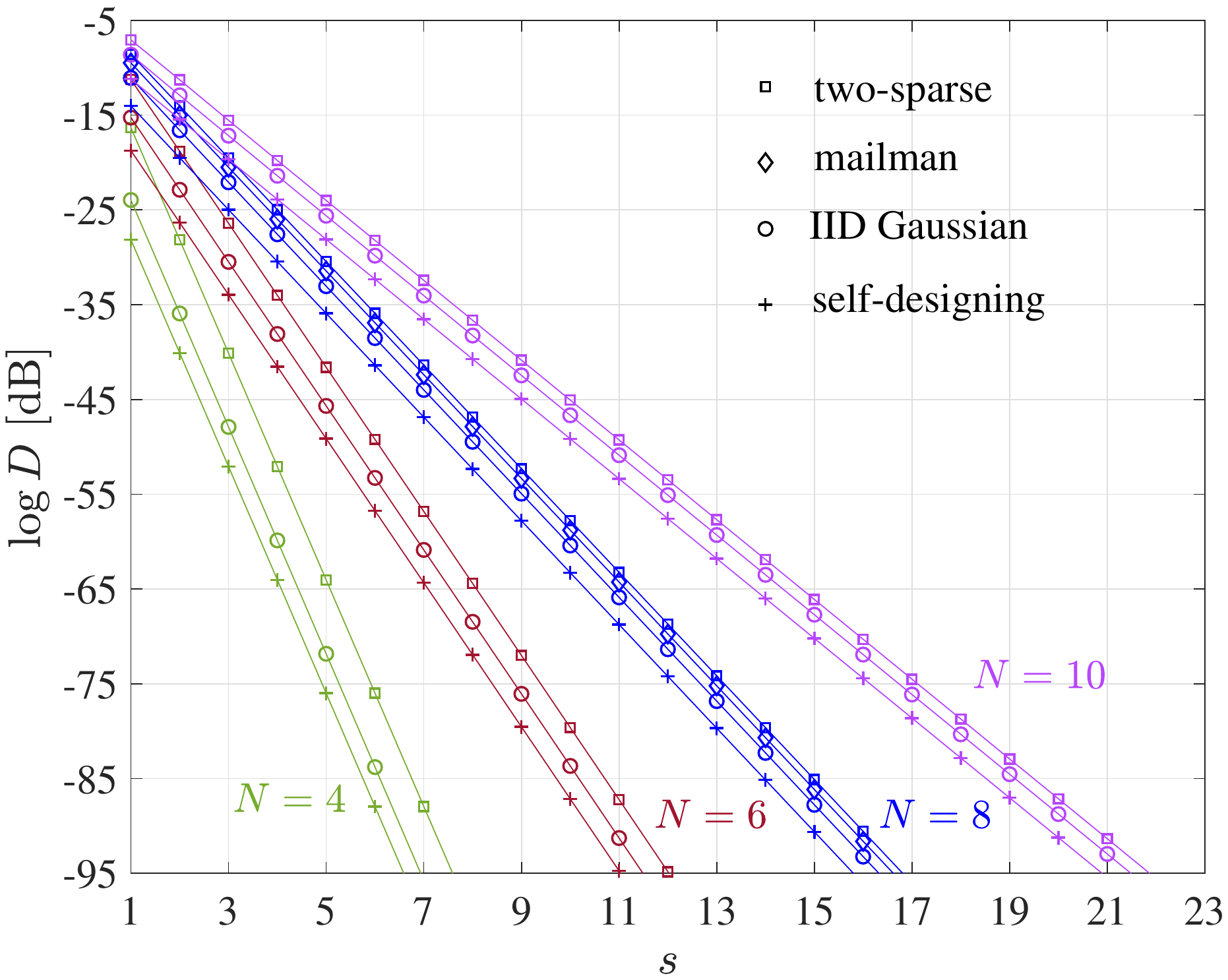,width=\columnwidth}
\caption{Average relative total squared error vs.\ total number of additions per column required for the wiring matrices and Gaussian IID target matrices of size $N\times 256$. Results are averaged over $10^6$ random matrix entries.
\label{k256}}
\end{figure}

For an IID Gaussian target matrix with $K=256$ columns, we compare various codebook matrices in Fig.~\ref{k256}. The wiring matrices are designed as described in Section~\ref{acclowbou}.
For all numbers of rows $N$ that are shown in the figure, self-designing codebooks perform best while two-sparse codebooks perform worst. The mailman codebook, which only exists for $N=8$, performs slightly worse than the IID Gaussian codebook. The choice of the codebook does not affect the slope of the curve, but is responsible for an offset shift in the average distortion.

\subsection{Number of Additions per Matrix Entry}

Due to their superior performance, we restrict the subsequent considerations to self-designing codebooks.
First, we address IID Gaussian target matrices. Here, the target matrix can be used to self-design the codebook matrix, as IID Gaussian codebooks work well. As in the previous subsections, we use multiple wiring matrices with $s_\ell=1$ for all of them.
Table~\ref{rect}
\begin{table}
\begin{center}
\caption{Number of additions per matrix entry that are required to reach the accuracy of 2-, 4-, 8-, 16-, and 24-bit signed integer arithmetic or better. Best values for given matrix width shown in boldface.
\label{rect}}
\begin{tabular}{||c||c|c|c|c|c|}
\hline
$\tilde C$ & 2 bit & 4 bit & 8 bit & 16 bit & 24 bit\\
\hline\hline
$4\times 256$ &0.5 &0.75 & 1.25 &  2.25 &3.25  \\
$5\times 256$ & 0.4 &0.8  & 1.2 &\bf 2.2&\bf 3.2\\
$6\times 256$ & 0.333 &0.667  & 1.167 &2.333&3.333\\
$7\times 256$ & 0.427 &0.714 &  \bf 1.143 &  2.286 & 3.286  \\
$8\times 256$ & 0.375 &0.625 &  1.25 &  2.25 & 3.375  \\
$9\times 256$ & 0.333 &0.667 &  1.222 & 2.333 &3.444   \\
$10\times 256$ & 0.3 & 0.6 &  1.2 &   2.3& 3.5    \\
$11\times 256$ & 0.364 & 0.636 &   1.182 &   2.364& 3.546    \\
$12\times 256$ & 0.333 & \bf 0.583 &  1.25 &   2.417& 3.583    \\
$13\times 256$ & 0.307 & 0.615 &  1.231 &   2.462& 3.615    \\
$14\times 256$ & \bf 0.286 & 0.643 &  1.214 &   2.5& 3.714    \\
$15\times 256$ & 0.333 & 0.6 &  1.267 &   2.533& 3.8   \\
\hline\hline
$5\times 1024$ & 0.4 & 0.6 &  \bf 1&  2 & 2.8  \\
$6\times 1024$ & 0.333 & 0.667 &  \bf 1&  1.833 &\bf 2.667  \\
$7\times 1024$ & 0.429 & 0.571 & \bf 1 & 1.857 & 2.714\\
$8 \times 1024$ & 0.375 &0.625 &  \bf 1 &   1.875 &2.75  \\
$9 \times 1024$ & 0.333 &0.556 &  \bf 1 &   1.889 &2.778  \\
$10 \times 1024$ &0.3  & \bf 0.5 &  \bf 1 &  1.9 &  2.7 \\
$11 \times 1024$ &0.273  & 0.546 &  \bf 1 &  \bf 1.818 &  2.727 \\
$12 \times 1024$ & 0.333& \bf 0.5 &  \bf 1 & 1.833 & 2.75  \\
$13 \times 1024$ & 0.308& 0.539 &  \bf 1 & 1.846 & 2.769  \\
$14 \times 1024$ & 0.286&  \bf 0.5 & \bf 1 & 1.857& 2.786\\
$15 \times 1024$ & 0.267&  0.533 & \bf 1 & 1.867& 2.8\\
$16 \times 1024$ & \bf 0.25 & \bf 0.5 & \bf 1 &1.875 &2.813 \\
$17 \times 1024$ & 0.294 & 0.529 & \bf 1 &1.882 &2.824 \\
\hline\hline
$7 \times 4096$ & 0.429& 0.571 &0.857&   1.571 & 2.286  \\
$8 \times 4096$ & 0.375& 0.5 &0.875&   1.625 & \bf 2.25  \\
$9 \times 4096$ & 0.333& 0.556 &0.889&   1.556 & 2.333  \\
$10 \times 4096$ & 0.3& 0.5 &0.9&   1.6 & 2.3  \\
$11 \times 4096$ & 0.273& 0.455 &0.818&   1.546 & 2.273  \\
$12\times 4096$ & 0.25 &0.5 & 0.833&  1.583 & 2.333    \\
$13\times 4096$ & 0.308 &0.462 & 0.846&   1.539 & 2.308    \\
$14\times 4096$ & 0.286& 0.427& 0.857 &1.571 & 2.286\\
$15\times 4096$ & 0.267& 0.467& 0.8 &\bf 1.533 & 2.333\\
$16\times 4096$ & 0.25& 0.438& 0.813 &1.563 &2.313 \\
$17\times 4096$ & 0.294& \bf 0.412& 0.824 &1.588 & 2.294\\
$18\times 4096$ & \bf 0.222& 0.444 & 0.833 &1.556 &2.333 \\
$19\times 4096$ & 0.263&  0.421 & \bf 0.790 &1.579 &2.316 \\
$20\times 4096$ & 0.25&  0.45 &  0.8 &1.55 &2.35 \\
\hline
\end{tabular}
\end{center}
\end{table}
shows the average number of additions per matrix entry to achieve at least the accuracy of standard signed integer arithmetic evaluated by \eqref{ssia}. For 8-bit accuracy, the code rate has little, for $K=1024$ even no impact.
For larger and lower accuracies, higher and lower code rates are favored, respectively.  
This trend is slightly distorted by quantization effects, but clear enough to be spotted. In comparison to the benchmark defined in Section~\ref{staofart}, the computational cost is reduced by 80 \%.

This behavior is explained as follows: 
There is the fixed computational cost of $2K$ additions for the codebook matrix. This cost is shared by the fewer rows, the larger is the code rate. This increases the computational burden per row (and also per matrix entry) for larger code rates.
For high accuracy, the wiring matrices clearly dominate the computational cost, so the computation of the codebook is secondary. Thus, higher rates are favored, as they generally result in lower distortions, see Fig.~\ref{totalerror}. For low accuracy, only few wiring matrices are needed and the relative cost of the codebook is more important. This shifts the optimum rate towards lower values.

Consider now a random matrix whose entries are uniform IID within $[0,1)$. If the same algorithm is used as before, the performance degrades significantly, as such a matrix is not a good codebook. Since all entries are positive, all codewords are constrained to a single orthant of the full space.
In order to circumvent this problem, we choose an auxiliary target matrix $\tilde \mT$ with Gaussian IID entries to self-design the codebook. In order to keep that codebook for all subsequent wiring, we use only a single wiring matrix and adapt the parameter $s_1$ in such a way as to reach the desired accuracy.
The results are shown in Table~\ref{uniform}.
\begin{table}
\begin{center}
\caption{Number of additions per matrix entry that are required to reach the accuracy of 2-, 4-, 8-, 16-, and 24-bit signed integer arithmetic or better.
\label{uniform}}
\begin{tabular}{||c||c|c|c|c|c|}
\hline
$\tilde C$ & 2 bit & 4 bit & 8 bit & 16 bit & 24 bit\\
\hline\hline
$8\times 256$ & 0.5 &0.75 &  1.25 &  2.375 & 3.5  \\
$10 \times 1024$ &0.4  &  0.6 &   1.1 &  2 &  2.9 \\
$12\times 4096$ & 0.333 &0.5 & 0.917&  1.75 & 2.5    \\
\hline
\end{tabular}
\end{center}
\end{table}
A comparison to Table~\ref{rect} shows that this procedure only causes a minor degradation, if any at all.

\section{Conclusions \& Outlook}
\label{conc}

Linear computation coding by means of codebook and wiring matrices is a powerful tool to reduce the computational cost of matrix-vector multiplications in deep neural networks.
The number of additions to achieve the accuracy of $q$-bit signed integer arithmetic is roughly given by
\begin{align}
\frac {2+q}{\log_2 K}.
\end{align}
This means no multiplication and only a single addition for 8-bit integer arithmetic and vectors with at least $K=1000$ dimensions.  

The idea of computation coding is not restricted to linear functions. Its direct application to multidimensional nonlinear functions promises even greater reductions in the computational cost of neural networks. We conjecture that neural networks need neither have precise weights nor be densely connected, if they are sufficiently deep. As in computation coding of linear functions, any quantization errors at intermediate layers and any level of sparsity may be compensated for with more layers and larger dimensions.

Fast matrix-vector products are also important for applications in other areas of signal processing, e.g., beamforming for wireless multiple-input multiple-output systems \cite{castaneda:20}, compressive sensing, numerical solvers for partial differential equations, etc. This opens up for many future research directions based on linear computation coding.

\section*{Acknowledgement}
The authors like to thank Veniamin Morgenshtern, Marc Reichenbach, and Hermann Schulz-Baldes for helpful discussions.

\section*{Appendix}

\subsection*{Average Distortion of Canonically Signed Digit Form}

For the canonically signed digit form, the set of reconstruction values is ${\cal R} = \{\pm 2^k: k\in \Z\}$. 
Let the target variable $t$ be uniformly distributed within $[0,1]$.
Thus, only reconstruction values for $k\le 0$ are used.

Consider the interval $[2^{k-1},2^k]$ with $k\le0$. The probability $p_k$ that the variable $t$ falls within that interval is equal to its width $w_k=2^{k-1}$.
The average mean-squared quantization error within that interval is well known to be $w_k^2/12$. Averaging over all intervals, the average mean-squared distortion for quantization with a single digit is given by
\begin{align}
 \frac1{12} \sum\limits_{k=-\infty}^0 p_k w_k^2 &= \frac 1{12} \sum\limits_{k=-\infty}^0 w_k^3 = \frac 1{12} \sum\limits_{k=-\infty}^0 2^{3k-3}\\
& = \frac1{96}  \sum\limits_{k=0}^\infty 8^{-k} = \frac 1{96} \cdot \frac 87=\frac1{3\cdot 28}.
\end{align}

By symmetry, the same considerations hold true, if $t$ is uniformly distributed within $[-1,0]$. Due to the signed digit representation, these considerations even extend to $t$ uniformly distributed within $[-1,+1]$.  

For representation by zero digits, $t\in[-1,+1]$ is quantized to 0 with average mean-squared distortion equal to $\frac13$. This is 28 times larger than for quantization with one digit.

The canonically signed digit representation is invariant to any scaling by a power of two. Thus, any further digit of representation also reduces the average mean-squared distortion by a factor of 28.

\subsection*{Additions Required for Binary Mailman Codebook}

Let $\mB_{N\times K}$ denote the binary mailman matrix with $N$ rows and $K$ columns.
It can be decomposed recursively as 
\begin{align}
\mB_{N\times K} = \left[\begin{array}{cc}
\mB_{(N-1)\times \frac K2} & \mB_{(N-1)\times \frac K2}\\
\mathbf 0_{1\times \frac K2} & \mathbf 1_{1\times \frac K2}
\end{array}\right].
\end{align}
Following \cite{liberty:09}, we decompose $h$ into its first half $h_1$ and its second half $h_2$ such that we get
\begin{align}
\label{recmm}
\mB_{N\times K} h = \left[
\begin{array}{c}
\mB_{(N-1)\times \frac K2} (h_1 + h_2)\\
\mathbf 1_{1\times \frac K2} h_2
\end{array}\right] .
\end{align}
Let $c(N)$ denote the number of additions to compute the product $\mB_{N\times K} h$.
The recursion \eqref{recmm}, implies
\begin{align}
c(N) & = \frac K2 + c(N-1)+ \frac K2 - 1\\
& < c(N-1) + 2^N.
\end{align}
These are $\frac K2$ additions for $h_1+h_2$, $c(N-1)$ additions for the matrix-vector product, and $\frac K2-1$ additions to sum the components of $h_2$.
Note that $\mB_{1\times 2}h = h_2$. Thus, $c(1)=0$ and $c(N) < 2^{N+1}=2K$.
 
\subsection*{Asymptotic Cumulative Distribution Function}

In order to show the convergence of the CDF of the squared angle error to the unit step function, recall the following limit holding for any positive $x$ and $u$
\begin{equation}
\lim\limits_{K\to\infty}\left(
1- \frac x{K^u}
\right)^K = \left\{
\begin{array}{ll}
0 & u<1\\
\exp(x) & u=1\\
1 & u >1
\end{array}
\right..
\end{equation}
The limiting behavior of ${\text P}_{a^2}(r)$ is, thus, decided by the scaling of $1-{\text P}_{\rho^2}(1-r)$ with respect to $K$. The critical scaling is $\frac 1K$.
Such a scaling implies a slope of $-1$ in doubly logarithmic scale.
Thus,
\begin{equation}
\lim\limits_{N\to\infty} \frac{\partial}{\partial (NR)}
\log_2 \left[1 - {\text P}_{\rho^2}(1-r)\right]  = -1.
\end{equation}
Explicit calculation of the derivative yields
\begin{equation}
\lim\limits_{N\to\infty}
\frac{
\int\limits^1_{1-r} (1-\xi)^{\frac {N-3}2 }\xi^{-\frac12} 
 \log_2\left(1-\xi\right)
{\text d}\xi.
}{2R
\int\limits^1_{1-r} (1-\xi)^{\frac {N-3}2} \xi^{-\frac12}  {\text d}\xi.
}=-1
\end{equation}
and saddle point integration gives \cite[Chapter 4]{merhav2010statistical}
\begin{equation}
\frac 1{2R} \log_2[1-(1- r)] =-1.
\end{equation}
This immediately leads to $r =4^{-R}$.

\bibliography{lit}
\bibliographystyle{IEEE}
\end{document}